\documentclass[preprint,showpacs,preprintnumbers,amsmath,amssymb]{revtex4}

\usepackage{graphicx}
\usepackage{dcolumn}
\usepackage{bm}

\begin{document}

\preprint{APS/123-QED}

\title{Cascaded Nondegenerate Four-Wave Mixing Technique for High-Power Single-Cycle Pulse Synthesis in the Visible and Ultraviolet Ranges}
\author{R. Weigand}
\affiliation{Departamento de \'Optica, Facultad de Ciencias F\'isicas, Universidad Complutense de Madrid, Ciudad Universitaria s/n, 28040 Madrid, Spain}
\email{weigand@fis.ucm.es}
\author{J. T. Mendon\c{c}a}
\affiliation{IPFN and CFIF, Instituto Superior T\'ecnico, Av. Rovisco Pais 1, 1049-001 Lisboa, Portugal}
\author{H. M. Crespo}
\affiliation{IFIMUP/IN - Institute of Nanoscience and Nanotechnology and Departamento de F\'isica, Faculdade de Ci\^encias, Universidade do Porto, R. do Campo Alegre 687, 4169-007 Porto, Portugal}
\email{hcrespo@fc.up.pt}   

\date{\today}

\begin{abstract}
We present a new technique to synthesize high-power single-cycle pulses in the visible and ultraviolet ranges by coherent superposition of a multiband octave-spanning spectrum obtained by highly-nondegenerate cascaded four-wave mixing of femtosecond pulses in bulk isotropic nonresonant media. The generation of coherent spectra spanning over two octaves in bandwidth is experimentally demonstrated using a thin fused silica slide. Full characterization of the intervening multicolored fields using frequency-resolved optical gating, where multiple cascaded orders have been measured simultaneously for the first time, supports the possibility of direct synthesis of near-single-cycle 2.2~fs visible-UV pulses without recurring to complex amplitude or phase control, which should enable many applications in science and technology.
\end{abstract}


\maketitle

Achieving efficient and reproducible generation of intense carrier-envelope phase (CEP) stabilized single-cycle light pulses in the visible and ultraviolet (UV) ranges is expected to have a strong impact in many fields, from time-resolved spectroscopy and coherent control of molecular dynamics~\cite{Kling2006} to the study of phase-sensitive nonlinear phenomena~\cite{Muecke2001} and the generation of isolated attosecond pulses~\cite{Sansone2006,Goulielmakis2008}.

To generate single-cycle pulses, a spectral bandwidth of the same order or larger than the central frequency (usually called an octave-spanning spectrum) is required, and the radiation must be coherent and properly phase-locked across this spectrum. For spectra consisting of multiple sidebands, the frequency of each sideband must also be an integer multiple of the interband spacing if one wishes to synthesize a train of identical CEP-stabilized pulses.

Several techniques have been proposed to generate high-power single-cycle pulses in the near-infrared (NIR) and visible ranges. Temporal compression of the broadband supercontinuum produced by self-phase modulation (SPM) in gas-filled hollow-fibers~\cite{Nisoli1997} has been extensively used as an efficient source of sub-terawatt few-cycle pulses. This technique, combined with adaptive compression, resulted in 1.4~GW pulses as short as 2.6~fs~\cite{Matsubara2007}, whereas passive compression using broadband chirped mirrors has recently allowed for the generation of CEP stabilized high-power 1.5-cycle pulses~\cite{Cavalieri2007}. A different but related approach based on self-compression by filamentation of ultrashort laser pulses in gases also holds promise for the generation of high-power CEP stabilized single-cycle pulses~\cite{Couairon2005,Goulielmakis2008_2}.

To our knowledge, the only experimental demonstrations of single-cycle optical pulse synthesis in the visible range~\cite{Shverdin2005,Chen2008} rely on molecular modulation driven in a gas by two independent nanosecond lasers~\cite{Harris1998}. This process generates multioctave spectra composed of very narrow Raman sidebands separated by the frequency difference between the driving lasers. By spatially dispersing seven collinear sidebands generated in Deuterium and adjusting their phases with a liquid crystal modulator, a train of $\sim 10^{6}$ single-cycle 1.6~fs pulses separated by 11~fs and with a peak power of 1~MW was obtained~\cite{Shverdin2005}. Very recently, trains of identical sub-cycle optical pulses with constant CEP within the train were synthesized from commensurate Raman sidebands produced in Hydrogen~\cite{Chen2008}. However, it remains difficult to isolate a single pulse from the train and the energy per pulse is very low.

In this paper we present and describe a new technique for synthesizing high-power CEP stabilized single-cycle pulses based on highly-nondegenerate cascaded four-wave mixing (CFWM). We will show that, by coherently superposing the multicolored phase-locked pulses produced by CFWM, a compact train of very short optical pulses can be obtained even without complex phase control.

The first demonstration of high-order nondegenerate CFWM in bulk transparent media~\cite{Crespo2000} showed that multiple broadband ultrashort pulses extending from the NIR to the UV can be efficiently generated by spatially and temporally overlapping two intense noncollinear femtosecond pulses with different frequencies in a nonresonant $\chi^{(3)}$ nonlinear medium. This phenomenon corresponds to coherent scattering of the incident pulses by the rapidly oscillating nonlinear polarization induced in the medium by the two-color pump field. More exactly, the two pump pulses with central frequencies $\omega_{0}$ and $\omega_{1}$ ($\omega_{1}>\omega_{0}$) drive the $\chi^{(3)}$ medium at the modulation (beat) frequency $\omega_m=\omega_1-\omega_0$, giving rise to multiple pulses with frequencies $\omega_n=\omega_0+n\omega_m$, where $n$ is the beam order and $n>1$ $(n<0)$ denotes frequency upconverted (downconverted) pulses. Simultaneous phase-matching is achieved by optimizing the pump interaction angle, and pulses with near-Gaussian beam profiles are emitted in a well-defined pattern where higher frequency shifts correspond to larger emission angles. The use of a 150-$\mu$m-thick glass slide as the nonlinear medium allowed for high conversion efficiencies (10-20$\%$ for Gaussian pumps) and the generation of frequency-upconverted beams up to the 11th order, while minimizing linear dispersion and competing nonlinear optical effects such as SPM, since the optical path length was simultaneously smaller than the Rayleigh range, the nonlinear length, and the dispersion length of the femtosecond pump pulses.

Nonresonant nondegenerate CFWM in the femtosecond regime was subsequently observed and demonstrated in other nonlinear media and spectral regions, from semiconductors pumped with mid-IR pulses~\cite{Chin2001}, to gases and plasmas pumped in the NIR to UV. In particular, Misoguti et al.~\cite{Misoguti2001} successfully extended CFWM to the vacuum-UV range (down to 160~nm) using a gas-filled hollow waveguide pumped with $\omega$ and $2\omega$ pulses from a Ti:sapphire laser. They obtained a conversion efficiency of 40\% in the first order pulse (a remarkably high value for a direct nonresonant $\chi^{(3)}$ process) while showing that CFWM allows for upconversion of light in very large energy steps (1.5~eV, or over 12000~cm$^{-1}$), unlike resonant processes such as anti-Stokes Raman scattering, being therefore uniquely suited for generating intense light beyond the UV.
These results evidence the universal nature of nonresonant CFWM processes, which can occur in quite different conditions and materials, provided the nonlinear medium is transparent to both the pump and the generated wavelengths, and the necessary phase-matching conditions can be met over a sufficiently broad range. They also confirm that frequency conversion using nondegenerate CFWM exhibits higher efficiencies and broader bandwidths than other schemes for generating light in the visible and the UV.

CFWM detains further characteristics of particular interest for pulse synthesis. In the important case of noncollinear phase-matching considered here, the new frequencies are already emitted at different propagation angles. This may obviate the need for dispersive elements such as the gratings or prisms used in SLM-based devices (see, e.g., Ref.~\cite{Shverdin2005}) in case additional amplitude and/or phase control is used prior to beam recombination to synthesize a given pulse shape. Also, and unlike in Raman-based processes, the central frequency of the total spectrum and the separation between sidebands can be freely adjusted by tuning the pump frequencies, so in principle it should always be possible to obtain the commensurate sidebands required to synthesize a train of identical pulses. Furthermore, the multiband spectra that can be generated by cascaded four-wave mixing comes to fill the gap between the usual supercontinuum spectra generated by self-phase modulation of ultrashort pulses in transparent media and the multiline spectra generated by molecular modulation in Raman media, which also has important consequences in the time domain. Since the corresponding octave-spanning spectrum is composed of broad bandwidths separated by the pump beat frequency, the pump energy can be channeled into a synthesized field consisting of only a few pulses or even a single pulse.

A deeper insight into pulse synthesis by coherent addition of a finite number of CFWM pulses can be obtained with a simplified theoretical model, based on coupled amplitude equations between the pumps and the generated sidebands derived from the wave equation in the slowly-varying envelope approximation and assuming a cubic nonlinearity~\cite{Crespo2001}. For two pump fields with complex amplitudes $\tilde E_0=E_0 e^{i \phi_0}$ and $\tilde E_1=E_1 e^{i \phi_1}$, frequencies $\omega_0$ and $\omega_1$, respectively, neglecting dispersion and pump depletion, and assuming perfect phase-matching and constant coupling coefficients, the solution for the complex amplitude $\tilde E_n$ of the field with frequency $\omega_n$ at the exit of a medium with length $L$ can be written as~\cite{Crespo2001,Weigand2be}
\begin{equation}
\label{1}
\tilde E_n(s)=i^n e^{i n \delta} \tilde E_0 J_n(s) - i^{n+1} e^{i(n-1)\delta} \tilde E_1 J_{n-1}(s),
\end{equation}
where $s=\gamma L$ is a nonlinear phase shift, $\delta=\phi_1-\phi_0$ is the initial phase difference between pumps, $J_n$ are the \textit{n}th order Bessel functions of the first kind, and the nonlinear coefficient is given by $\gamma=3 \omega_0 \chi^{(3)} E_0 E_1/(2 n_0 c)$, with $n_0$ the linear refractive index at frequency $\omega_0$ and $c$ the speed of light in vacuum.  Equation~(\ref{1}) predicts well-defined phase relationships between the several CFWM orders, thus establishing mutual coherence between the beams. The total field in the time domain is then
\begin{equation}
\label{2}
\textstyle{E(t)=\text{Re} \big[\sum_n \tilde E_n(s) e^{-i(\omega_0+n \omega_m)t} \big].}
\end{equation}
From Eq.~\eqref{2} and using the Jacobi-Anger identity, $\sum_n J_n(z) \exp(i n \theta) i^{n}= \exp(i z \cos \theta)$, we obtain
\begin{eqnarray}
\label{3}
\nonumber E(t)=Re \big[ e^{i\phi_0}(E_0 e^{-i \omega_0 t}+ E_1 e^{i \delta}e^{-i \omega_1 t}) \\
\times e^{-i s \cos(\omega_m t - \delta)} \big].
\end{eqnarray}
This result is similar to the FM solution by Harris \textit{et al.}~\cite{Harris1998} for coherent addition of Raman sidebands in the approximation of negligible dispersion and limited modulation bandwidth, with the important difference that in CFWM the synthesized pulse envelope strongly depends on the initial phase difference $\delta$. In molecular modulation, the detuning between the driving frequency and the Raman transition dictates the preparation of a phased or an anti-phased molecular state, the latter resulting in a negative effective nonlinear coefficient $\gamma$ for which the synthesized pulses could have an approximately negative chirp and hence could be further compressed by simple propagation in a normally dispersive medium~\cite{Harris1998}. In the present case of CFWM, the sign of $\gamma$ is independent of the pump field (and usually positive for most optically transparent $\chi^{(3)}$ media), but the chirp of the synthesized pulses can nevertheless be controlled by adjusting $\delta$ alone. From Eq.~(\ref{3}) we see that $\phi_0$ only affects the CEP fo the pulse train, defined with respect to the two-color beat envelope, whereas $\delta$ is associated with net temporal shifts and also determines the pulse chirp, as given by the oscillating nonlinear phase term. In particular, when $\delta=\pm\pi/2$, the synthesized pulses can have a negative chirp. In practice, a given phase difference can be introduced by slightly adjusting the delay between the two phase-locked pump pulses without significantly affecting their temporal overlap. Also, if the pump frequencies are commensurate, this will result in the generation of identical pulses that are CEP-stabilized within each train (although their CEP may change from shot-to-shot), since even though each CFWM step is not self-CEP-stabilized, the total recombined field will be. The corresponding pulse envelopes will also be identical from shot-to-shot, provided that only the relative phase difference $\delta$ remains constant, which is a far less stringent requirement than the need of CEP stabilized pump pulses.

\begin{figure}[htb]
\begin{center}
\includegraphics[width=10 cm]{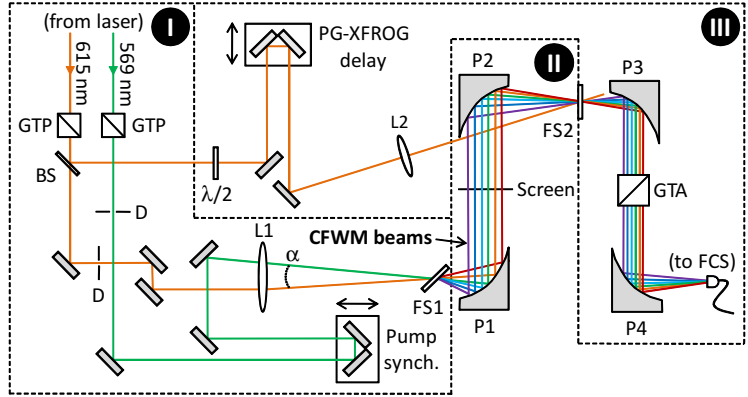} 
\caption{\label{Setup}(color online) Experimental setup: (I) generation of CFWM pulses; (II) pulse recombination and synthesis; (III) pulse characterization by PG-XFROG (see text for details).}
\end{center}
\end{figure}

To demonstrate the feasibility of the proposed technique, we implemented the experimental setup in Fig.~\ref{Setup}, which comprises generation of the CFWM beams (I), pulse synthesis (II), and diagnostics (III).
As in previous work~\cite{Crespo2000}, two horizontally polarized visible femtosecond pulses from a dual-wavelength (orange: $\lambda_0$=615~nm, $\sim$80~fs, 2~mJ; green: $\lambda_1$=569~nm, $\sim $60~fs, 200~$\mu$J) 10~Hz dye laser amplifier are used as pumps. In this laser system, the green pump beam is directly (optically) derived from the orange pump beam: a small portion of the orange beam is first used to generate supercontinuum in a cell filled with deuterated water, and then the green portion of this supercontinuum is amplified using green laser dyes. Even though supercontinuum generation and laser action are coherent processes, the relative phase between the two pulses is not actively locked and can fluctuate due to normal thermal and mechanical perturbations in the system. The system is not tunable, since the two colors are fixed by the gain bandwidths of the respective laser dyes, but the two visible pump wavelengths spaced by some tens of nanometers are nevertheless quite adequate to generate CFWM spectra composed of many, closely spaced, broad bandwidths in the visible and UV, which are useful for synthesizing short pulse trains in these important spectral regions.

The pulses, which are commensurate to within 0.2\% (well within their $\sim$~5~nm bandwidths), are passed through Glan-Taylor polarizers (GTP; 100000:1 extinction ratio) and diaphragms (D) prior to focusing with a best-form lens (L1), at an external noncollinear interaction angle $\alpha\sim 3^{\circ}$, onto a 150-$\mu$m-thick fused-silica slide (FS1) oriented at $45^\circ$. The tilt angle is important to prevent total internal reflection of the generated higher order CFWM beams and also contributes to equalizing the optical path differences between the angularly dispersed beams. Fused-silica was chosen as the nonlinear medium because its transparency in the UV (down to 180~nm) and low dispersion favor noncollinearly phase-matched CFWM over broader bandwidths than standard optical glass~\cite{Crespo2008_1}.

\begin{figure}[b]
\begin{center}
\includegraphics[width=10 cm]{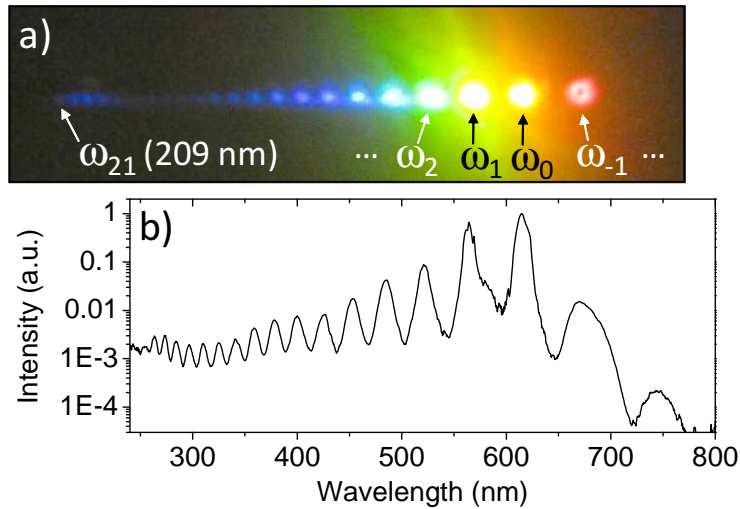} 
\caption{\label{Directimage}(color online) (a) Direct (unfiltered) image of the fan of multicolored CFWM pulses as seen projected on a phosphor-coated paper screen (the dark arrows denote the pump beams). (b) Corresponding two-octave spectrum measured at the focal plane of mirror P2.}
\end{center}
\end{figure}

At the entrance plane of slide FS1, the synchronized orange and green pulses have energies of 32 and 38~$\mu$J, respectively, near-transform-limited durations, and similar intensities on the order of $2 \times 10^{12}$~W/cm$^2$, generating two frequency-downshifted and a fan of frequency-upshifted CFWM beams up to the 20th order, or 209~nm, as shown in Fig.~\ref{Directimage}(a). To our knowledge, this is the highest order process ever generated by CFWM of femtosecond pulses. The total measured energy in the cascaded beams (excluding orders 0 and 1) is $>6$~$\mu$J, so approximately 10\% of the incident energy is transferred to the newly generated frequencies.
Pulse recombination and coherent addition of the cascaded orders was implemented with a low dispersion setup [Fig.~\ref{Setup}(II)] based on two carefully aligned $\lambda$/8 Aluminum-coated off-axis parabolic mirrors  P1 (f=2.54~cm) and P2 (f=5.08~cm) to first collimate and then focus the angularly separated multicolored pulses into a small ($\sim100$~$\mu$m) white-light spot.
Figure~\ref{Directimage}(b) shows the corresponding spectrum measured with an intensity calibrated UV-NIR (200-1100~nm) fiber-coupled spectrometer, where only 15 frequency upconverted orders could be detected due to the limited bandwidth of the mirrors.
The spectrum nevertheless spans over two octaves and could potentially support (with proper amplitude and phase control) sub-cycle pulse durations.

Since CFWM results from a fast nonresonant nonlinearity of electronic origin, simultaneous phase-matching ensures that all pulses are phase-locked when exiting slide FS1. The short (30~cm) air path from FS1 up to the focal plane of mirror P2 contributes only a small amount of second-order dispersion with negligible higher-order dispersion, and the surface figure of the mirrors minimizes phase distortions; an apertured screen placed between mirrors P1 and P2 allows us to eliminate most of the residual pumps by transmitting only their more intense central portions. Hence, we expect a train of few-cycle pulses to be Fourier-synthesized at the focus of P2.

\begin{figure}[htb]
\begin{center}
\includegraphics[width=10 cm]{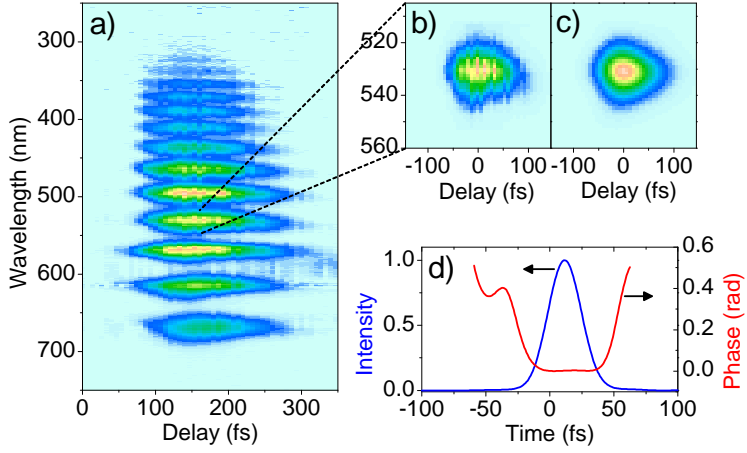} 
\caption{\label{PGXFROG}(color online) (a) Measured PG-XFROG trace of the synthesized field. (b) Measured and (c) retrieved PG-XFROG traces of the first frequency-upconverted CFWM pulse, and (d) corresponding intensity and phase in the time domain.}
\end{center}
\end{figure}

\begin{figure}[t]
\begin{center}
\includegraphics[width=10 cm]{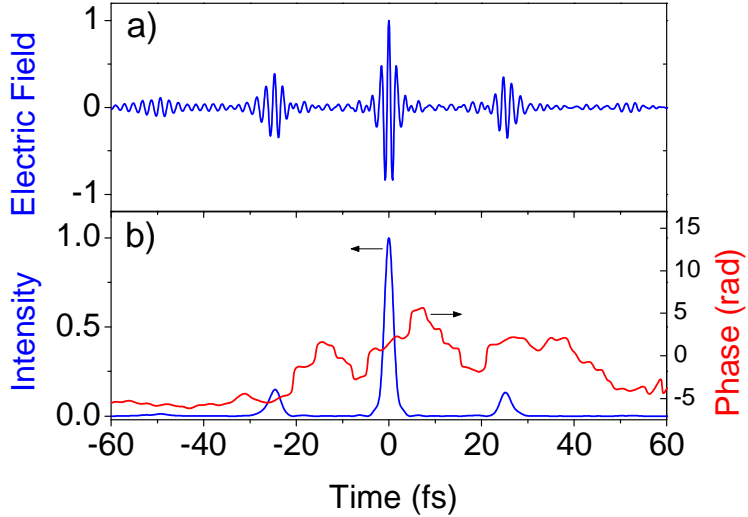} 
\caption{\label{Totalfield}(color online) (a) Normalized electric field and (b) intensity and phase of the pulses obtained by coherent addition of the retrieved electric fields of the 13 gated CFWM pulses. The main peak is a 1.3-cycle, transform-limited 2.2~fs pulse.}
\end{center}
\end{figure}

Temporal characterization of the pulses [Fig.~\ref{Setup}(III)] was performed with a broadband polarization-gated cross-correlation frequency-resolved optical gating (PG-XFROG) technique. A portion of the orange pulses was extracted with a 50/50 beamsplitter (BS), polarization rotated by 45$^{\circ}$ with a half-wave plate and delayed with a motorized stage to serve as gate by inducing birefringence in a second fused silica slide FS2 (similar to FS1) placed at the focal plane of mirror P2, where recombination of the CFWM beams and pulse synthesis take place. The reference and the generation arms have practically the same (precompensated) dispersion. Measurement of the gated CFWM field required a second set of parabolic mirrors, where mirror P3 (f=2.54~cm) collimated and directed the gated beams through a crossed UV Glan-Thompson analyzer (GTA), and mirror P4 (f=5.08~cm) launched the transmitted pulses into a 400-$\mu$m-core intensity calibrated fiber-coupled spectrometer (FCS). In this low-dispersion setup, the high degree of pump polarization provided by the Glan-Taylor polarizers was directly transferred to the CFWM beams, unlike in standard polarization-gated setups where signals must first cross a dispersive polarizing element prior to gating. To increase the signal-to-noise ratio, PG-XFROG traces were registered by averaging ten gated spectra for each time delay (6.7~fs time steps). A typical PG-XFROG trace is shown in Fig.~\ref{PGXFROG}(a) clearly revealing multiply synchronized, near-chirpless pulses produced by nondegenerate CFWM of the transform-limited pumps.
To our knowledge this is the first time-resolved measurement of multiple CFWM processes, comprising 13 simultaneously gated pulses. From the measured XFROG trace, we see that near-dispersionless focus-to-focus imaging between slides FS1 and FS2 was achieved. To test the sensitivity of the setup, a 450~$\mu$m glass plate was placed between mirrors P1 and P2, resulting in clearly observable delays between the consecutive CFWM pulses, in agrement with the known dispersion of the plate.
Because this particular laser system is not CEP stabilized, the pump phases $\phi_0$ and $\phi_1=\phi_0+\delta$ will be different in each shot, although fluctuations in the relative phase difference $\delta$ are mostly due to thermal drifts and should evolve at a relatively slow rate. Absolute phase changes should not affect the envelopes of the individual CFWM pulses, but possible changes in $\delta$ will nevertheless cause the coherently synthesized pulses to be shifted in time within the train and their shape to change from shot-to-shot, resulting in smearing of the temporal structure in the acquired PG-XFROG traces. More importantly however, the long duration of the available gate pulses prevents direct observation of fine temporal structure in the traces and also hinders the direct retrieval of a complete trace starting from initial noise. The complete electric field of each of the 13 gated pulses can nevertheless be retrieved unambiguously, as illustrated in Figs.~\ref{PGXFROG}(b), (c) and (d) for the first frequency-upconverted order for which we obtained a 30.6~fs (FWHM) transform-limited pulse with a time-bandwidth product (TBP) of 0.37 and a XFROG error of 0.012 (128$\times$128 grid size). The transform-limited orange reference pulses (87~fs FWHM) were characterized with a standard FROG technique, implemented by removing slide FS1 from the XFROG setup. All retrievals were performed using commercial software (\textit{Femtosoft FROG}), and TBPs between 3.3 and 4.7 with XFROG errors between 0.01 and 0.02 were consistently obtained for all but the last three (and much less intense) UV orders. The gated CFWM field was therefore fully characterized apart from the unknown but fixed (assuming a constant $\delta$) phase differences between the cascaded pulses. By setting these phases to zero (performed by introducing small, fixed time delays between each cascaded pulse) we obtain a synthesized field composed of a main transform-limited 2.2~fs (FWHM) pulse - less than 1.3 optical cycles at the center wavelength of 513~nm - and two smaller side pulses at a distance of 25~fs (the pump beat period), as shown in Fig.~\ref{Totalfield}.
Since most of the energy in the total field is concentrated in a single main pulse, CEP stable pulses can be synthesized even for non-commensurate pumps pulses, provided they are CEP stabilized. The energy in the pulse is $\sim$~5~$\mu$J, corresponding to a peak power of $\sim$~2~GW.

We would like to point out that even if beam recombination and pulse synthesis take place at a small region within the focal plane of mirror P2, it should be possible to synthesize a collimated pulse by using a diffraction grating placed at the focal plane of mirror P2 to cancel the angular separation of the CFWM beams, although at the expense of bandwidth since gratings are limited to one octave.

In conclusion, we have presented and demonstrated a technique to efficiently Fourier-synthesize high-power few- to single-cycle pulses in the visible-UV range from the multiband coherent spectra generated by cascaded nondegenerate four-wave mixing of femtosecond pulses in bulk nonresonant $\chi^{(3)}$ media. Broadband PG-XFROG measurements were performed for the multiple CFWM orders for the first time, showing the possibility of synthesizing 2.2~fs 1.3-cycle pulses with multi-gigawatt peak powers without significant manipulation of the intermediate cascaded beams. The use of phase-stable transform-limited pump and gate pulses with durations in the 25-30~fs range, such as those provided by a CEP-stabilized Ti:sapphire laser amplifier (possibly coupled to an optical parametric amplifier or hollow-fiber compressor to obtain two phase-locked pump pulses with a given frequency separation) will significantly improve the generation and measurement (through the shorter gate pulses) of near-single-cycle trains of identical pulses and CEP-stabilized isolated single-cycle pulses using this technique, which should have a strong impact in many fields of science and technology.

\begin{acknowledgments}
Financial support by the Access to Research Infrastructures activity in the Sixth Framework Programme of the EU (contract RII3-CT-2003-506350, Laserlab Europe) for conducting research at the Laboratoire d'Optique Appliqu\'ee, ENSTA - \'Ecole Polytechnique, France is gratefully acknowledged. We thank J. Etchepare and G. Mourou for insightful discussions, and A. dos Santos for technical assistance.
\end{acknowledgments}


\end{document}